\documentclass[%
 aip,
 amsmath,amssymb,
 reprint,%
]{revtex4-1}

\usepackage{graphicx}% Include figure files
\usepackage{dcolumn}% Align table columns on decimal point
\usepackage{bm}% bold math
\usepackage[utf8]{inputenc}
\usepackage[T1]{fontenc}
\usepackage{mathptmx}
\usepackage{etoolbox}
\usepackage{xcolor}
\usepackage{xr}
\usepackage{filecontents}
\makeatletter
\def\@email#1#2{%
 \endgroup
 \patchcmd{\titleblock@produce}
  {\frontmatter@RRAPformat}
  {\frontmatter@RRAPformat{\produce@RRAP{*#1\href{mailto:#2}{#2}}}\frontmatter@RRAPformat}
  {}{}
}%
\makeatother
\begin{document}

\preprint{AIP/123-QED}

\title[Photoelectron Spectrosocopy and Circular Dichroism of an Open-Shell Organometallic Camphor Complex]{Photoelectron Spectroscopy and Circular Dichroism of an Open-Shell Organometallic Camphor Complex}

\author{Viktoria Brandt}
\author{Michele Pugini}
 \email{dstemer@fhi-berlin.mpg.de}
\affiliation{Fritz-Haber-Institut der Max-Planck-Gesellschaft, Faradayweg 4-6, 14195 Berlin, Germany
}%

\author{Nikolas Kaltsoyannis}
\affiliation{Department of Chemistry, The University of Manchester, Oxford Road, Manchester M13 9PL, England	
}%

\author{Gustavo Garcia}
\affiliation{Synchrotron SOLEIL, l’Orme des merisiers, Départementale 128, St. Aubin 91190, France
}%

\author{Ivan Powis}
\affiliation{School of Chemistry, The University of Nottingham, University Park, Nottingham, NG7 2RD, England
}%

\author{Laurent Nahon}
\affiliation{Synchrotron SOLEIL, l’Orme des merisiers, Départementale 128, St. Aubin 91190, France
}%

\author{Dominik~Stemer$^{\;\text{1}, *}$}

\date{\today}% It is always \today, today,
             %  but any date may be explicitly specified

\begin{abstract}
 We present an investigation of one-photon valence-shell photoelectron spectroscopy and photoelectron circular dichroism (PECD) for the chiral molecule (1R,4R)-3-(heptafluorobutyryl)-(+)-camphor (HFC) and its europium complex Eu(III) tris[3-(heptafluorobutyryl)-(1R,4R)-camphorate] (Eu-HFC$_{3}$), the latter of which constitutes the heaviest organometallic molecule for which PECD has yet been measured. We discuss the role of keto-enol tautomerism in HFC, both as a free molecule and complexed in Eu-HFC$_{3}$. PECD is a uniquely sensitive probe of molecular chirality and structure such as absolute configuration, conformation, isomerisation, and substitution, and as such is in principle well suited to unambiguously resolving tautomers; however modeling remains challenging. For small organic molecules, theory is generally capable of accounting for experimentally measured PECD asymmetries, but significantly poorer agreement is typically achieved for the case of large open-shell systems. Here, we report PECD asymmetries ranging up to $\sim8\%$ for HFC and $\sim7\%$ for Eu-HFC$_{3}$, of similar magnitude to those reported previously for smaller isolated chiral molecules, indicating that PECD remains a practical experimental technique for the study of large, complicated chiral systems.

\end{abstract}

\maketitle

\section{Introduction}

Photoelectron spectroscopy (PES) is powerful method for the investigation of molecular electronic structure. Nowadays, when the molecule of interest is chiral, PES can be enhanced by the complementary technique of photoelectron circular dichroism (PECD). Chiral light-matter interactions have long been leveraged to resolve the physically and chemically indistinguishable enantiomers of various chiral molecules. The differential absorption of circularly polarized light in the infrared and UV domains has proven a particularly accessible means of probing the electronic and vibrational structure of chiral molecules. However, because such absorption-based circular dichroism (CD) relies on the interaction between electronic and much weaker magnetic dipole transitions, the magnitude of the observed effects tends to be low (typically 10$^{-3}$--10$^{-1}$~\%). In contrast, chiral asymmetry in the photoelectron angular distribution arises purely from electric-dipole interactions, and correspondingly the measured effects are larger (typically 0.5 -- 10's~\%), even for randomly oriented molecules.\cite{powis_PECDchapter_2008, Nahon_Review_2015, Sparling_PECDrev_2025}

PECD, which manifests as a forward-backward asymmetry in the photoelectron flux emitted along the light-propagation axis in photoionization of chiral molecules by circularly polarized light, was first predicted theoretically in 1976 by Ritchie.\cite{ritchie_PECDtheory1, ritchie_PECDtheory2} The first experimental demonstrations of this effect were published in the early 2000s.\cite{bowering_PECD_2001, Garcia_PECD_2003} In subsequent years, a number of exploratory PECD studies were performed on a range of terpenes, including camphor and fenchone,\cite{hergenhahn_PECD_2004, nahon_valencecamphorPECD1_2006, nahon_isomerPECD_2016, powis_valencecamphorPECD2_2008, C1sPECD_ulrich, nahon_isomerPECD_2016} as well as several smaller systems\cite{Nahon_Review_2015} thereby establishing PECD as a general phenomenon in photoionization of chiral molecules. The dependence of PECD on the final photoelectron continuum states (and thus electron kinetic energy) is clear from core-level photoionization investigations, e.g. of the C 1s orbital in camphor.\cite{C1sPECD_ulrich, hergenhahn_PECD_2004} As a 1s orbital is spherical (achiral) and highly localized, PECD asymmetry measured upon photoionization of such orbitals clearly indicates that the outgoing photoelectrons must sample the intrinsic chirality of the full  molecular potential via scattering processes. Although camphor and fenchone exhibit very similar PECD in C1s photoionization within the kinetic-energy range of 1--15~eV,\cite{C1sPECD_ulrich, hergenhahn_PECD_2004} photoionization of their highest-occupied molecular orbitals (HOMOs) reveals remarkable differences, despite the fact that in both cases the HOMO is strongly localized on the molecule's respective carbonyl group.\cite{Garcia_PECD_2003, powis_valencecamphorPECD2_2008, nahon_isomerPECD_2016} 

This extraordinary sensitivity to minor differences in the initial state of the molecules, as well as to the molecules' overall geometry and valence electronic structure, makes PECD an attractive technique capable of providing new insights that may aid in the interpretation of complicated photoelectron spectra. Indeed, experiment and theory both clearly indicate that PECD is much more sensitive to minor changes in a molecule's geometric or electronic structure than either of the more commonly measured photoionization observables; cross section and anisotropy parameter ($\beta$).\cite{PECDtheory_Stener_2004, powis_PECDchapter_2008,hartweg_condensation-PECD_2021, dupont_pecd-conformer_2022}

However, despite PECD's exceptional sensitivity, its application as a practical analytical method demands reasonable agreement between theory and experiment. A major challenge for theoreticians, in turn, is the accurate description of electron continuum states which, together with the neutral electronic ground states, are needed to determine the photoionization matrix elements that give rise to PECD.\cite{stener_terpeneTheory_2006, demekhin_single_2011,decleva_continuum_2022} For the case of small to moderately sized organic molecules without significant electron correlation, this information is generally accessible and thus the agreement between PECD theory and experiment is sufficient to enable the practical use of PECD, for example in the resolution of different conformers that are indistinguishable in the angle-integrated PE spectrum.\cite{hadidi_conformer-dependentPECD_2021,tia_alaPECD2_2014, turchini_alaninolPECD_2009, Daly_epich_2011} Even for more complicated molecules, such as closed-shell organometallic complexes, agreement between theory and experiment can be nearly quantitative for well-characterized ionization channels.\cite{catone_resonant_2012} However, for open-shell molecules (which exhibit stronger electron correlation), or for ionization channels involving deeper valence states, and therefore not well approximated by one-electron excitations, PECD remains challenging to model, and agreement with experiment has generally been poorer.\cite{catone_photoelectron_2013, darquie_Ru-AcAc-PECD_2021} 

Improvements in modeling will require comparison to experimental data. In the case of chiral organometallic complexes, the study of which could benefit from the unique capabilities of PECD, such data remains sparse. These molecules are of broad scientific interest both practical applications and tests of fundamental physics. They are regularly employed as chiral catalysts and shift reagents for nuclear magnetic resonance,\cite{aspinall_chiral_2002} and as efficient emitters of circularly polarized electroluminescence.\cite{zinna_CPL_2015} At a more basic level, such molecules are interesting due to their capacity to filter polarized electrons via elastic scattering at levels measurable in the laboratory,\cite{mayer_experimentalverification_1995, mayer_electrondichroism_1996} and are useful candidates for experiments seeking to quantify miniscule energetic differences between enantiomers due to parity violation.\cite{quack_high-resolution_2008, darquie_progress_2010}

\begin{figure}[h!]
	\includegraphics[width=\linewidth]{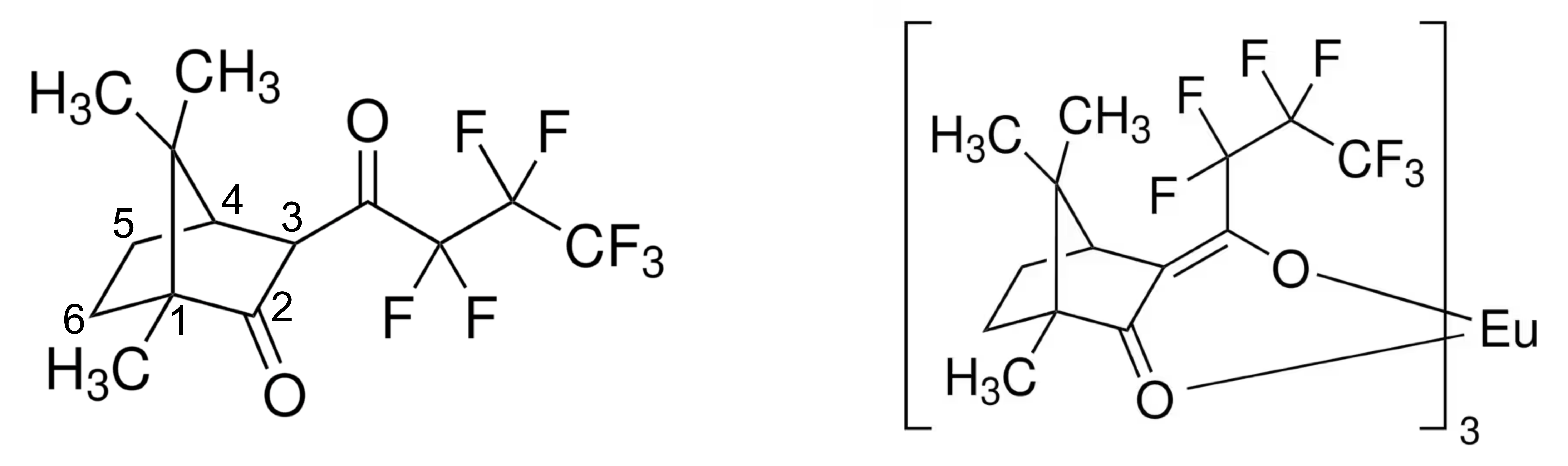}
	\caption{Commercially provided molecular structures of (1R,4R)-3-(heptafluorobutyryl)-(+)-camphor (HFC, left) and Eu(III) tris[3-(heptafluorobutyryl)-(1R,4R)-camphorate] (Eu-HFC$_{3}$, right).} 
	\label{fig:Structures}
\end{figure}

%Chiral organometallic molecules are of interest for a range of practical applications including chiral catalysis,\cite{aspinall_chiral_2002} as shift reagents for nuclear magnetic resonance,\cite{aspinall_chiral_2002} and as efficient emitters of circularly polarized electroluminescence.\cite{zinna_CPL_2015} At a more fundamental level, such molecules are interesting due to their capacity to filter polarized electrons via elastic scattering at levels measurable in the laboratory,\cite{mayer_experimentalverification_1995, mayer_electrondichroism_1996} and are useful candidates for experiments seeking to quantify miniscule energetic differences between enantiomers due to parity violation.\cite{quack_high-resolution_2008, darquie_progress_2010}
%PECD and can provide a very sensitive probe not just of the enantiomeric excess but also more elusive structural properties such as conformation.\cite{Turchini_2017} 
% PECD manifests as a forward-backward asymmetry in the photoelectron flux emitted along the light-propagation axis in photoionization of a chiral molecule by circularly polarized light.\cite{powis_PECDchapter_2008, Nahon_Review_2015, Sparling_PECDrev_2025}
Here, we report on our recent PES and PECD experiments involving gas-phase samples of (1R,4R)-3-(heptafluorobutyryl)-(+)-camphor (hereafter abbreviated HFC, see Fig.~\ref{fig:Structures}, left) and its complex with europium, Eu(III) tris[3-(heptafluorobutyryl)-(1R,4R)-camphorate] (hereafter Eu-HFC$_{3}$, see Fig.~\ref{fig:Structures}, right). HFC is a camphor derivative that differs from camphor via the introduction of a large, flexible, electronegative tail at the C$_3$ position, adjacent to the camphor carbonyl group. Eu-HFC$_{3}$ is composed of three HFC molecules coordinated to a central trivalent europium cation. We note that Eu-HFC$_{3}$ is at present the heaviest organometallic molecule to have been probed using PECD (m/z = 1194), and therefore serves as a useful candidate for assessing PECD's relevance as an analytical tool to study heavy molecules with multiple chiral centers.

\section{Methods}

\subsection{Experimental}

All measurements were carried out at the DESIRS VUV beamline at Synchrotron SOLEIL.\cite{nahon_desirs_2012} To prepare a sufficiently dense gas-phase target using HFC (resp. Eu-HFC$_{3}$), 2~mL of the liquid sample (both molecules were obtained from Santa Cruz Biochemicals and used without further treatment) were placed into a stainless steel oven, which was subsequently heated to 55$^\circ$~C (resp. 195$^\circ$~C). The vapor was expanded through a 100~$\mu$m pinhole, heated to 75$^\circ$~C (resp. 215$^\circ$~C), with 1~bar He carrier gas. The supersonic expansion was skimmed using two skimmers with diameters of 1.5~mm and 2~mm, respectively. The molecular beam was perpendicularly crossed with a VUV photon beam in the center of a i$^2$PEPICO DELICIOUS III PEPICO spectrometer coupling a velocity-map-imaging spectrometer on the electron side with a modified Wiley-McLaren 3D momentum imaging spectrometer on the ion side.\cite{garcia_delicious_2013} Photoelectron spectroscopy was performed using circularly polarized light in the photon-energy range of 9--13~eV. For PECD, measurements were made by collecting image pairs with left- and right-handed circularly polarized light, switching polarization at $\sim15$ minute intervals. These were combined following an established protocol to obtain symmetric and antisymmetric image for subsequent analysis.\cite{nahon_valencecamphorPECD1_2006} The resulting mass-selected photoelectron images were analyzed using the pBasex inversion procedure.\cite{Garcia_igor_2004}

Within the electric-dipole approximation, the normalized photoelectron angular distribution resulting from one-photon ionization of a randomly oriented molecule by circularly polarized radiation may be described as

\begin{equation}
    I^{\{p\}}(\theta) = [1+b^{\{p\}}_{1}P_{1}(cos\theta)+b^{\{p\}}_{2}P_{2}(cos\theta)],
\end{equation}

where $P_{n}$ represents the $n$th-order Legendre polynomial.\cite{powis_PECDchapter_2008} For circular polarizations, $\theta$ is the angle of photoemission with respect to the vector defined by photon propagation. The index $p$ describes the helicity of the light; $p = \pm1$ for left-handed (right-handed) circularly polarized light and $p=0$ for linear polarizations. Since $P_{1}(\cos\theta))$ expands simply as
$\cos\theta$, the chiral asymmetry parameters, $b^{\{\pm1\}}_{1}$, determine the magnitude of forward-backward asymmetry. From underlying symmetry considerations one obtains $b^{\{+1\}}_{1} = -b^{\{-1\}}_{1}$, but for linear polarizations $b^{\{0\}}_{1}$ is necessarily zero. For circular polarizations the $P_2$ Legendre coefficient, $b^{\{p\}}_{2}$, is symmetric so that $b^{\{+1\}}_{2}$ = $b^{\{-1\}}_{2}$, and for linear polarization $b^{\{0\}}_{2} \equiv \beta $, the conventional photoelectron anisotropy parameter. 

The magnitude of the measured chiral forward-backward asymmetry is generally defined as $2b^{\{p\}}_{1}$. In this study we use the shorthand $b_{1} = b^{+1}_{1}$, and $PECD=2b^{\{+1\}}_{1}$ when discussing derived numerical values of the chiral asymmetry.

\subsection{Computational\label{sec:compute}}

Quantum chemical calculations were carried out as follows. The Gaussian 16 software package, revision C.01, was used for all density functional theory calculations.\cite{g16} The hybrid density functional approximation, PBE0,\cite{PBE0_theory_Ernzerhof,PBE0_theory_Adamo} was used with Grimme’s D3\cite{D3_theory_Grimme} and the Becke-Johnson damping parameters for dispersion corrections.\cite{DFT_Becke_Johnson1, DFT_Becke_Johnson2, DFT_Becke_Johnson3} Dunning’s correlation consistent basis sets of polarized triple-$\zeta$ quality were employed for H, C, O, and F.\cite{basisset_Dunning4,basisset_Dunning1,basisset_Dunning2,basisset_Dunning3} A Stuttgart-Bonn relativistic effective core potential was used for Eu (28 electrons), with the associated segmented valence basis sets.\cite{relativistic_theory1,relativistic_theory2,relativistic_theory3} Eu(III) has the electronic configuration [Xe]4f$^6$ and hence Eu-HFC$_{3}$ was computed as a spin-unrestricted septet. Spin contamination was minimal, with <S$^2$> = 12.04 for Eu-HFC$_{3}$. Default settings were used for the SCF and geometry optimizations (which were performed without symmetry constraints), and analysis of the harmonic vibrational frequencies confirmed the optimized geometries as energetic minima. 

Ionization energies for the  PBE0/cc-pVTZ optimized  HFC geometries were calculated using the outer valence Green's function (OVGF) method\cite{Zakrzewski1996} implemented in Gaussian 16 and the non-Dyson third-order algebraic-diagrammatic construction scheme (IP-ADC(3[4+]))\cite{Dempwolff2020, Dempwolff2022} implemented in Q-Chem 5.4.\cite{QChem5} Both OVGF and IP-ADC(3[4+]) methods address one-hole (1h) independent-electron ionization processes through third-order many-body perturbation theory, thereby incorporating a treatment for electron correlation effects.

Neither the OVGF nor the ADC(3) calculations were feasible for the much larger Eu-HFC$_3$ complex. Ionization energies for this were instead estimated by the $\Delta E_{SCF}$ method determining the energy difference between the neutral and cation states. 

\section{Results}

\begin{figure*}[ht]
	\includegraphics[width=\textwidth]{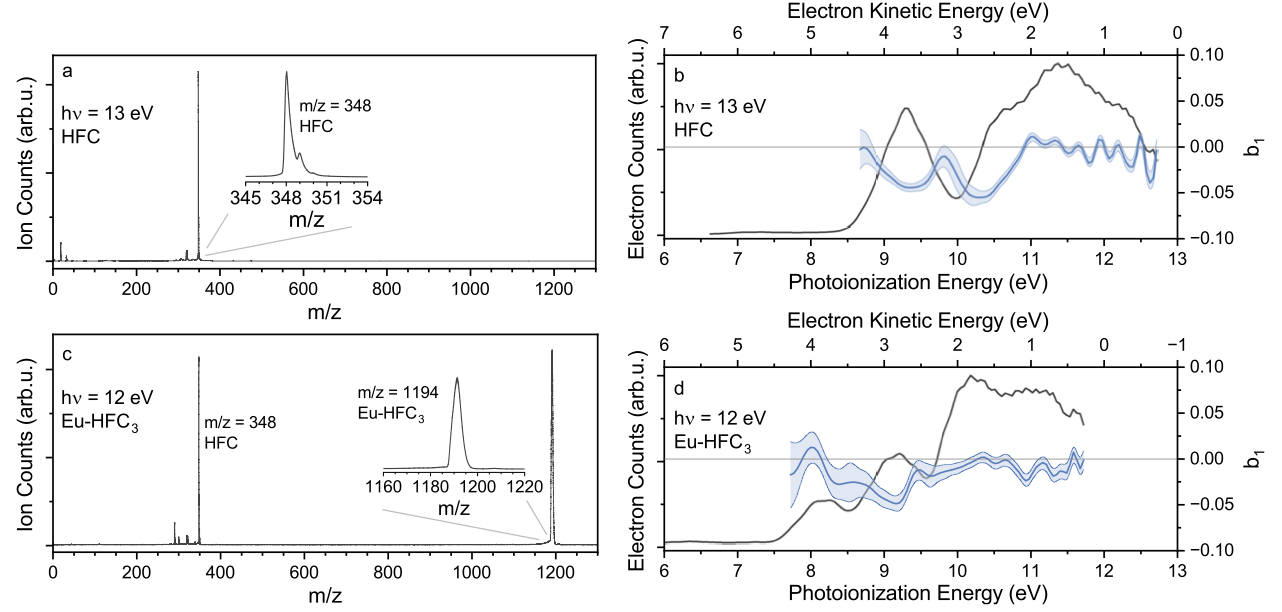}
	\caption{Representative photoion mass time-of-flight spectra (left) and corresponding photoelectron spectra collected in coincidence with the parent ion peaks (right) for (a,b) HFC and (c,d) Eu-HFC$_{3}$. The top row HFC spectra (a,b) were recorded with 13~eV photon energy; the bottom row Eu(HFC)$_3$ (c,d) measurements used 12~eV photons. Expanded views of the parent ion peaks are presented in (a) and (c). Photoelectron circular dichroism traces, showing $b^{\{+1\}}_{1}$ values derived from the photoelectron velocity map images, are presented as blue traces in (b) and (d).}
	\label{fig:PES}
\end{figure*}

Representative time-of-flight (TOF) spectra obtained with the highest photon energy used for both HFC and Eu-HFC$_{3}$ (13~eV and 12~eV, respectively) are displayed in Fig.~\ref{fig:PES}a \& c, with regions of interest highlighted in the insets. In the HFC TOF spectrum, we observed clear signatures of the parent ion at m/z~=~348. Fragmentation of HFC is also apparent at this photon energy, revealed via the small feature at m/z~=~320. However, in contrast to previous experiments with camphor, the parent ion remains dominant at all photon energies used,\cite{nahon_isomerPECD_2016} indicating a greater stability of HFC upon photoionization. For Eu-HFC$_{3}$ the parent ion peak at m/z~=~1194 is also clear. In this case, we did not observe any clear signatures of fragmentation. The size of Eu-HFC$_{3}$ (339 vibrational modes) evidently constitutes a greater heat-bath capable of holding the excess energy of ionization at the photon energies we examined, thereby stabilizing the system against fragmentation. The strong peak at m/z~=~348 originates from unreacted ligand, and is not a fragmentation product of Eu-HFC$_{3}$. This is clearly established by considering the PES obtained in coincidence with the ions, which is identical to that for HFC. For the subsequent analysis of the PE spectra of HFC and Eu-HFC$_{3}$, we include only electrons collected in coincidence with the respective parent ion masses.

The PE spectrum of HFC (Fig.~\ref{fig:PES}b) reveals one well-separated feature with a vertical ionization energy (VIE) of approximately 9.4~eV, as well as a more complicated envelope of overlapping features beginning at slightly higher IE, but with a distinct shoulder suggesting a second orbital ionization at $\sim 10.5$ eV.  As seen in Fig.~\ref{fig:PES}b, the corresponding HFC PECD exhibits clear negative maxima centered at the IEs corresponding to these first two features in the PES. At higher ionization energy, as the apparent density of spectroscopic states increases, the PECD trends to zero.

The measured narrow valence-band spectrum of Eu-HFC$_{3}$ (Fig.~\ref{fig:PES}d) is similar to that measured for HFC, with the notable difference being the presence of a new lower-energy feature with VIE~$\approx$~8.2~eV. Once again, two clear PECD extrema are seen at the IEs corresponding to the first two PE features. Similarly to HFC, we find that the higher IE features do not yield any notable PECD.

\begin{figure}[ht]
	\includegraphics[width=0.5\textwidth]{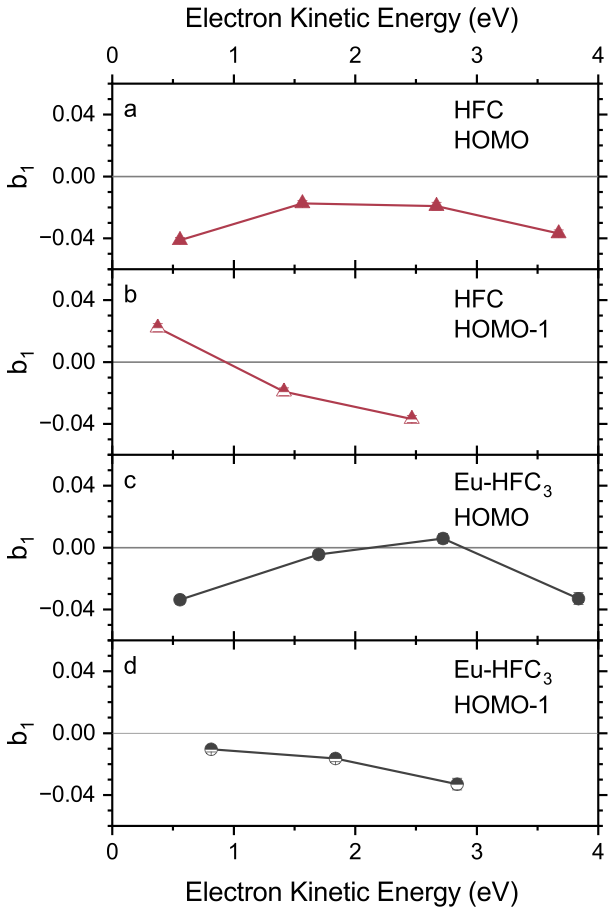}
	\caption{Photoelectron circular dichroism measured as a function of photoelectron kinetic energy for the two highest occupied molecular orbitals of (a,b) HFC and (c,d) Eu-HFC$_{3}$. Statistical error bars are included, but are generally smaller than the markers.} 
	\label{fig:PECD}
\end{figure}

To facilitate a more direct comparison between the PECD measured for each of the above-assigned PE features for HFC and Eu-HFC$_{3}$, we averaged the PECD $b_1$ values across the full-width at half maximum of each peak as a function of photon energy. The summarized data is presented in Fig.~\ref{fig:PECD}. Beginning with the HOMO of HFC, we find clear PECD across the entire kinetic-energy range probed, with the magnitude of $b_{1}$ varying between -0.014 and -0.042. The case of the HFC HOMO-1 is quite different, with a maximum measured value of $b_{1}$ of -0.037 at 2.6~eV KE, a slight decrease in magnitude at 1.6~eV KE, and a sign reversal with $b_{1}$ = 0.022 for the lowest-KE electrons studied.

For Eu-HFC$_{3}$ the measured PECD magnitude accompanying the first two PES peaks are in general lower, with the HOMO exhibiting a maximum magnitude for $b_{1}$ of -0.034 for 0.6~eV photoelectrons, near zero PECD for electrons with KE of 1.7~eV and 2.7~eV, and negative PECD once more for 3.7~eV photoelectrons. Eu-HFC$_{3}$'s HOMO-1 exhibited a maximum measured value of $b_{1}$ of -0.033 for 2.8~eV photoelectrons, with PECD magnitude decreasing for lower KE electrons.

\section{Discussion}

\subsection{HFC}

Following an automated search\cite{Kim2013} of the conformational space of (1R,4R)-3-heptafluorobutyryl-(+)-camphor, we identified ten potential conformers of interest, which are reported in the PubChem database.\cite{PubChemHFC} Upon further examination, the heptafluorobutyryl substitutions at the tetrahedrally coordinated C$_3$ camphor atom immediately divide into two subclasses; half occupy the endo position, half the exo position. The C$_3$ atom thus becomes asymmetrically substituted with either R or S configurations possible, creating a second independent chiral center.\footnote{The unsubstituted camphor structure already has two stereogenic centers at the C$_1$ and C$_4$ atomic sites. However, because of the rigid bicyclic ring structure, the configuration at the latter is automatically determined by that of the former. Hence the R/S character of C$_4$ does not vary independently of C$_1$, and only a single enantiomer pair results in the unsubstituted camphor.}

To aid interpretation of the photoelectron spectra, we re-optimized these ten PubChem HFC conformer structures\cite{PubChemHFC} using the PBE0 hybrid functional with empirical dispersion corrections (as discussed in section \ref{sec:compute}) to obtain more accurate SCF energies. These PBE0/cc-pVTZ results are listed in Table~\ref{SI-tab:confEn}. 

Apart from the endo- or exo- nature of the substitution site, these conformers differ principally in the rotamer conformations adopted in the CF$_2\cdot$CF$_2\cdot$CF$_3$ tail grouping, the bicyclic structure of the camphor moiety being relatively rigid. The estimated OVGF/cc-pVTZ ionization energies for the outermost orbitals (listed in Table~\ref{SI-tab:confIE}) show very little dependence on the specific conformer structure. This invariance is perhaps not surprising as these outer orbitals localize around the more conformationally rigid camphor grouping rather than the floppier, electronegative  CF$_2\cdot$CF$_2\cdot$CF$_3$ tail (Fig.~\ref{SI-fig:HFC6mos}). Indeed the HFC valence PES (Fig.~\ref{fig:PES}b) bears a strong similarity to that of camphor,\cite{powis_valencecamphorPECD2_2008} fenchone,\cite{nahon_isomerPECD_2016} and bromocamphor\cite{bowering_PECD_2001} measured at comparable photon energies. For each of these molecules, the distinct HOMO band for each of these molecules may be attributed predominantly to the localized carbonyl orbital with prominent O lone-pair character. The greater VIE of the HFC HOMO compared to these other terpenes reflects the high electronegativity of the C$_{3}$ tail group.\cite{pollmann_terpeneUPS_1997}

\begin{figure*}[ht]
	\includegraphics[width=0.886\textwidth]{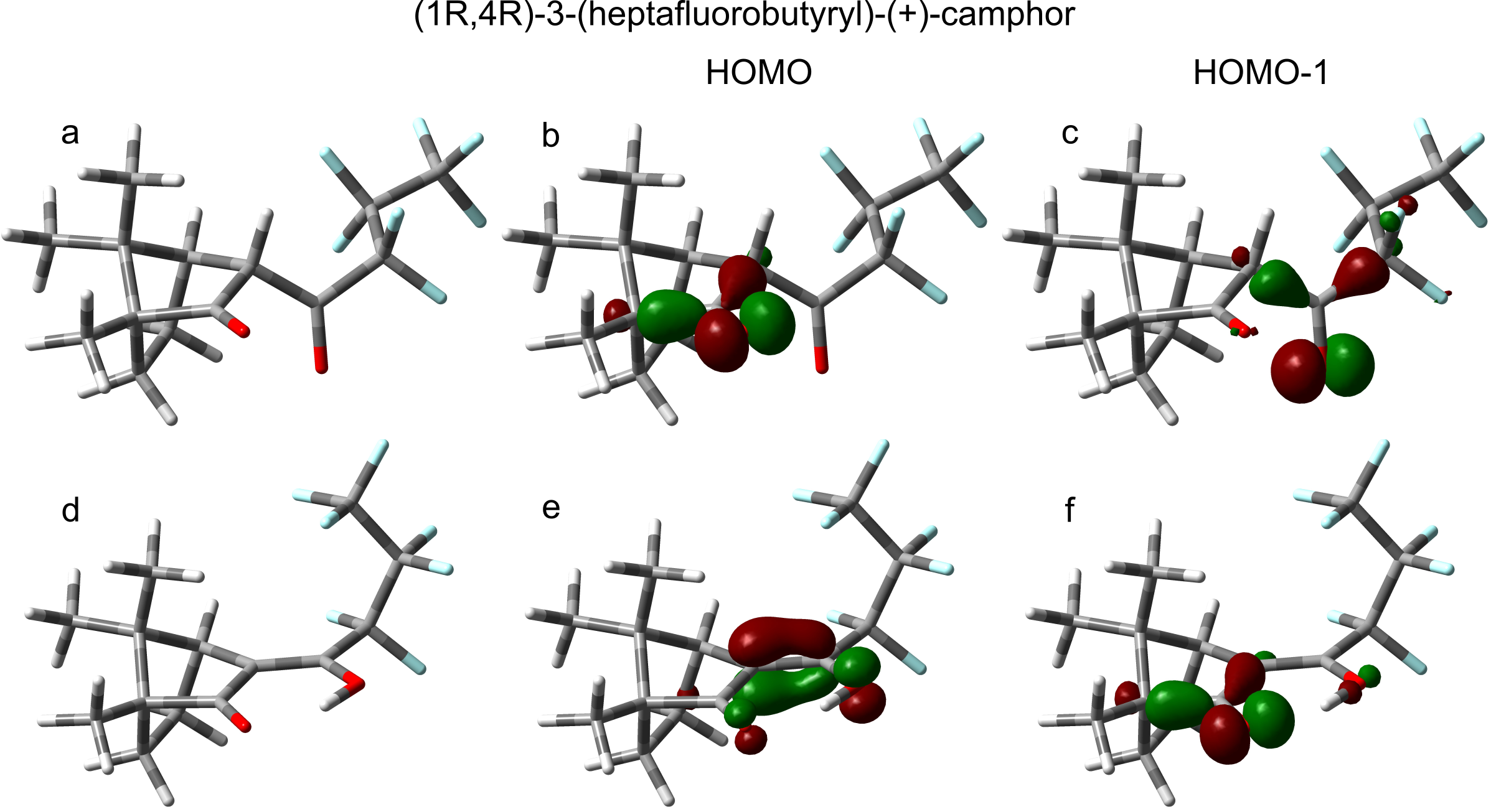}
	\caption{Molecular structures of the calculated lowest-energy conformers of the keto (a) and enol (d) tautomers of (1R,4R)-3-(heptafluorobutyryl)-(+)-camphor, along with visualizations of the two highest occupied molecular orbitals for each tautomer (b: keto HOMO; c: keto HOMO-1; e: enol HOMO; f: enol HOMO-1). For the molecular orbital visualizations, green and red represent the phases of the wavefunction with an isovalue of 0.1. All molecular orbitals shown are representative of the neutral species.}
	\label{fig:ligandMOs}
\end{figure*}

The molecular structure of the lowest-computed energy conformer (keto conformer \#9 in Table~\ref{SI-tab:confEn}), is shown in Fig.~\ref{fig:ligandMOs}a. The OVGF ionization energies for this conformer are 9.33, 10.56, 10.76, 11.30 and 11.55 eV  for the HOMO -- HOMO-4 orbitals respectively. The vertical ionization energies for the HOMO and HOMO-1, in particular, are in excellent accord with the visible peak positions in the PES and PECD spectra (Fig.~\ref{fig:PES}b). We thus conclude that these calculations provide a convincing assignment for at least the first two distinct photoelectron bands. Although our IP-ADC(3[4+]) calculations were more limited in overall scope, these results (Table~\ref{SI-tab:ADC-IP}) are in good agreement with the OVGF results.

Each of the above-discussed HFCs structure is of the diketone form illustrated in Fig.~\ref{fig:Structures}, which is how HFC is generally presented in vendor catalogs and online databases. However, we note that this is neither the only possible structure, nor even necessarily the most energetically stable one. In an experimental and computational vibrational circular dichroism (VCD) study of the closely related $\beta$-diketone molecule 3-(trifluotoacetyl)-camphor (TFA),\cite{merten_TFC_2010} the possibility of keto-enol tautomerism was considered. In the enolization process, a proton migrates from the C$_3$ atom creating a C$=$C double bond at that point and an adjacent OH group which may then H-bond toward the second C$_2=$O carbonyl group, resulting in the formation of a hydroxyketone group (the analogous HFC enol structure is shown in Fig.~\ref{fig:ligandMOs}d). Using DFT calculations with a polarizable continuum model, the authors found that, for the case of solution-phase TFA in chloroform, the enol form of the molecule was $\sim$19 kJ mol$^{-1}$ more stable than the keto forms. The experimental VCD spectrum matched well with that calculated for the most stable enol tautomer, and differed substantially from the calculated keto spectrum, providing further evidence for the dominance of the keto form of the molecule in solution.

%Tautomerisation has consequences for CD measurement, as the proton migration from the C$_3$ atom leaves it with a planar $sp^2$ hybridization. Consequently, it is no longer an asymmetrically substituted stereogenic center and the enol forms are no longer diastereomeric but just retain the C$_1$ enantiomerism.

While these solution-phase results for TFA may not directly apply to gas phase TFC, they strongly recommended the consideration of additional HFC structures. As such, we have extended our calculations to include HFC enol structures. Specifically, we generated trial structures starting from the two lowest-energy keto forms (keto conformers \#9 and \#1 in Table~\ref{SI-tab:confEn}) to define the CF$_2\cdot$CF$_2\cdot$CF$_3$ tail conformation and re-optimised with a PBE0/cc-pVTZ calculation. We found that the enol is predicted to be $\sim$30 kj mol$^{-1}$ more stable than the keto forms, suggesting that it would dominate any equilibrium mixture. While we have not investigated the HFC enol tail conformers as extensively as was done for the keto forms, our experience with the latter suggests that this will not significantly influence the broad conclusions to be drawn.
%There are, to our knowledge, no references to the diasteromeric character of the keto forms of HFC in the literature or suppliers' catalogs, from which one might infer that the diastereomers are not resolvable. The destruction of the second stereogenic centre in the HFC enol form, leaving the single camphor-like RS enantiomer, may well rationalise this, either because of a static dominance of the enol form, or by creating a racemization mechanism proceeding via dynamic keto$\leftrightarrow$enol interconversions.     

Without solvation effects, this dramatic stabilization in the gas-phase HFC enol must be attributed to formation of a six-membered H-bonded ring structure along with the increased conjugation in the enol (Figure~\ref{fig:ligandMOs}d). These structural changes are correspondingly apparent in the orbitals, notably those located near the hydroxyketone group. In particular, the enol HFC HOMO spans O$=$C-C$=$C conjugation, while the HOMO-1 looks remarkably similar to the keto HOMO, with clear localization on the C$_2=$O carbonyl group (compare Fig,~\ref{fig:ligandMOs}b~\&~e). Both our OVGF and ADC(3) ionization energy calculations (Tables \ref{SI-tab:confIE} and \ref{SI-tab:IP-ADC}) show that the VIE of the enol HOMO is strongly shifted to $\sim$8.9 eV, approximately 0.4 eV below the keto HOMO ionization.

It is apparent that the predicted enol HOMO vertical ionization energy is in rather worse agreement with the experimental PE spectrum (Fig.~\ref{fig:PES}b) than were the earlier keto HFC predictions. This creates a paradox since the foregoing arguments for a dominant role of the enol form seem to be in conflict with the experimental PES evidence. One may speculate that a high keto-enol interconversion energetic barrier could explain the apparent preference of the keto form in the PES data as a consequence of kinetics, in spite of the strongly favorable equilibrium energetics. Ultimately the PECD data measured may provide an ideal means to clarify the diastereomeric character of gas-phase HFC, and thereby to provide insights into the keto-enol equilibrium in this system that are not directly accessible via the PES. However, despite the strong HFC PECD signal reported here, the use of the PECD data in this analytic capacity will require reliable, high-quality theoretical modeling, which is beyond the scope of the present study.

\subsection{Eu-HFC$_{3}$}

\begin{figure*}[ht]
	\includegraphics[width=\textwidth]{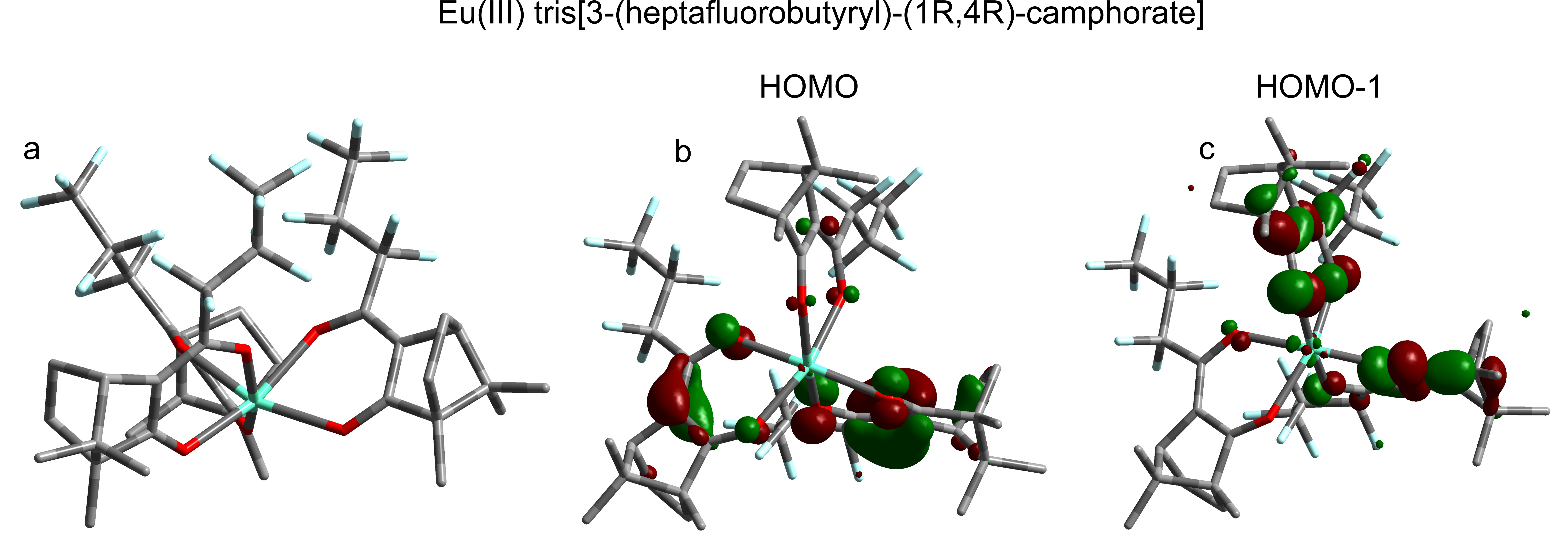}
	\caption{Optimized molecular structure of Eu-HFC$_{3}$ (a) with molecular orbitals visualized for the two highest occupied molecules orbitals (b: HOMO; c: HOMO-1). For the molecular orbital visualizations, green and red represent the phases of the wavefunction with an isovalue of 0.05. All molecular orbitals shown are representative of the neutral species. The differing perspective in (a) was chosen to highlight the helical ligand packing arrangement, while (b) and (c) utilize a different perspective to better illustrate the MOs.}
	\label{fig:ComplexMOs}
\end{figure*}

We optimized the structure of Eu-HFC$_{3}$ beginning with the bidentate ligand binding geometry suggested by Whitesides and Lewis.\cite{whitesides_EuComplex} The resulting structure is shown in Fig.~\ref{fig:ComplexMOs}a. We note that this structure is nearly C$_{3}$ symmetric, and as such, may be expected to exhibit helical, or P--M, chirality. For this type of chirality, the sign of the "twist" in the ligand packing configuration distinguishes two enantiomers, even for the case when the ligands themselves are achiral. This is similar, but not identical to, the case of the D$_{3}$-symmetric Ru(acac)${3}$, for which PECD has been previously reported.\cite{darquie_Ru-AcAc-PECD_2021} The class of X-HFC$_{3}$ molecules, where X is a trivalent lanthanide ion, thus provide new opportunities to study the interplay between intrinsic chirality, arising from the handedness of the ligands themselves, and structural chirality, dependent on the arrangement of the ligands. In the present case, we have no indication that P--M isomerism was resolved during the Eu-HFC$_{3}$ synthesis, and as such the molecules studied are likely racemic in this respect.

We note that the photoelectrons coincident with Eu-HFC$_{3}$'s parent ion comprise a similar spectrum (Fig.~\ref{fig:PES}d) to that measured for HFC, with the notable difference being the presence of a new lower-energy feature with IE~$\approx$~8.2~eV. This PE feature is clearly attributable to Eu-HFC$_{3}$'s HOMO, and our calculations reveal that the corresponding electron density for this orbital is primarily delocalized across the ring defined by the Eu-binding diketonate groups of the ligands (see Fig.~\ref{fig:ComplexMOs}b). This is different from the case of Ru(acac)$_{3}$,\cite{darquie_Ru-AcAc-PECD_2021} for which the HOMO was strongly metal localized, and reflects the lower stability of the occupied metal valence orbitals for transition metal complexes generally. The Eu-HFC$_{3}$ HOMO exhibits three-fold energetic degeneracy (i.e. there are two more energetically equivalent orbitals to that shown in Fig.~\ref{fig:ComplexMOs}b, see Table~\ref{SI-tab:Orbital_Energies}, right column, orbitals 1--3), reflecting the molecule's approximate C$_{3}$ symmetry and indicating equal electron density on each of the three coordinating ligands. 

Spectroscopically, the HOMO-1 can clearly be resolved as a second well-separated peak, centered around 9.2~eV IE. The electron density for this orbital is also delocalized across the ligand diketonate group, but exhibits substantially more oxygen lone-pair character compared to the HOMO, as well as minor contribution from the Eu~4f orbitals (Fig.~\ref{fig:ComplexMOs}c). As with the HOMO, this MO also exhibits three-fold energetic degeneracy (see Table~\ref{SI-tab:Orbital_Energies}, right column, orbitals 4--6). While it was not practical to calculate the valence VIEs for Eu-HFC$_{3}$, we used the so-called $\Delta$-SCF method (i.e. the energetic difference between the neutral and cationic molecular ground state) to estimate the HOMO VIE at 7.695~eV. This is somewhat of an underestimate based on comparison with the PE spectrum. Nevertheless, our calculated orbital energetics revealed a $\sim$~0.95~eV energy difference between the HOMO and HOMO-1 of Eu-HFC$_{3}$ (Table~\ref{SI-tab:Orbital_Energies}), which agrees well with the measured data and supports these feature assignments. The third PE feature, centered near IE~$\approx$~10~eV, is formed by the overlap of the HOMO-3 and HOMO-4, with more strongly bound MOs combining to form a higher-energy plateau beginning at 10.5~eV.

The conformational landscape of Eu-HFC$_{3}$ is difficult to define given the many possible rotations of the ligand CF$_2\cdot$CF$_2\cdot$CF$_3$ tail grouping. However, given the likely ligand packing arrangement presented in Fig.~\ref{fig:ComplexMOs}a, we note that some cooperative steric stabilization of the ligand tails may be expected. Given the bidentate binding geometry of the ligands to the Eu(III) ion, it appears probable that the ligands adopt a quasi-enol geometry, with the Eu(III) serving as the sixth member of a stable ring structure involving the ligand diketonate group. Moreover, we note that the HOMO and HOMO-1 of Eu-HFC$_{3}$ more closely resemble those of enol HFC than keto HFC (compare Figs.~\ref{fig:ComplexMOs}b~\&~c to Figs.~\ref{fig:ligandMOs}e~\&~f), with the HOMO delocalized across the diketonate group and the HOMO-1 more localized on the O lone pairs. The measured VIE of the Eu-HFC$_{3}$ HOMO more closely resembles the calculated values for enol HFC than for keto HFC, and the VIE of the Eu-HFC$_{3}$ HOMO-1 is nearly aligned with the VIE of the keto-HFC HOMO, seemingly reflecting the similarities to the enol HFC HOMO-1. While this comparison is not exact, both the calculated VIEs and MOs of the enol HFC more readily fit to the measured Eu-HFC$_{3}$ data, strongly suggesting "locking-in" of enol structure for HFC ligands upon coordination to Eu(III). Here again, the PECD data could be the definitive piece of evidence. Unfortunately, PECD calculations for such large organometallic molecules remain exceptionally difficult at the moment, and their implementation here is not yet practical.

We note that the PECD measured for the HOMO and HOMO-1 of Eu-HFC$_{3}$ and HFC, when compared (Fig.~\ref{fig:PECD}), share qualitative similarities in terms of sign and kinetic-energy trends. As the chirality of Eu-HFC$_{3}$ stems from the chirality of its individual HFC ligands, it may be tempting to ascribe these similarities to this shared chirality between the molecules. However, given that the HOMO and HOMO-1 of each molecule differ substantially, and that our PES data and calculations suggest different dominant tautomers for free and coordianted HFC, this apparent consistency seems likely to be simply a coincidence. Even if the MOs were very similar, complexes of chiral molecules are generally clearly distinguishable from their monomers in terms of PECD.\cite{nahon_dimerPECD_2010, hartweg_condensation-PECD_2021, Powis_glycidol_clusters_2014}

\section{Conclusions}

We have recorded photoelectron spectra and PECD spectra for the gas-phase ligand HFC and its Eu(III) complex, Eu(HFC)$_3$. We utilized DFT calculations to examine the relative stabilities and ionization energies of different conformations of the free HFC molecule. Despite the considerable flexibility of the CF$_2\cdot$CF$_2\cdot$CF$_3$ tail grouping, a rather consistent picture emerges with the ionization energies of the outer orbitals, which are localized around the more rigid camphor structure, being in good agreement with the experimental PES results. The role of keto-enol tautomerism in the gas phase HFC is discussed, with calculations indicating that the enol form is the more stable form, and so likely to be dominant in an equilibrium sample. Nevertheless, the agreement between the calculated enol HOMO ionization energy and the experimental photoelectron spectrum is worse, so that the PES evidence tends rather to support a predominance of the keto form.

One significant consequence of the keto-enol conundrum is that the keto structures are expected to be diastereomers (due to the two independent chiral centers). While cis-trans isomerism is technically possible for the enol structure across the enol double bond, our calculations indicate that the configuration with the tail hydroxyl group hydrogen bonded to the camphor carbonyl is clearly energetically preferred, meaning that enol HFC is in practice likely to be enantiomeric. This has obvious consequences for PECD measurements, and in principle the measured PECD should be well-placed to resolve keto vs. enol HFC. However, such calculations near the edge of feasibility using current methods, and are beyond the scope of the present paper.

For the case of Eu-HFC$_{3}$, the calculated HOMO and HOMO-1 orbitals are almost solely ligand-localized, with the Eu(III) 4f orbitals playing a minor role at most. Our lowest-energy calculated molecular structure of Eu-HFC$_{3}$ suggests that the coordinated HFC ligands adopt an enol-like structure, with the Eu(III) ion serving as one part of a stable six-membered ring involving the ligand diketonate groups. The measured PES data further support this picture. We note that the approximate C$_{3}$ symmetry of Eu-HFC$_{3}$ provides an opportunity to study the interplay between the intrinsic chirality of the individual HFC ligands and the structural P--M chirality of the complex, determined by the nature of the ligand packing. PECD would be well suited to resolve these contributions, but additional challenges remain in modeling such large molecules.

It is encouraging that the magnitude of PECD asymmetry observed for Eu-HFC$_{3}$ is comparable to that of HFC despite the differences in size and complexity between the systems. This suggests that PECD may be a viable analytical technique even for large organometallic systems. HFC is a large molecule as far as current calculations of PECD are concerned, and Eu-HFC$_{3}$ is the largest organometallic molecule for which PECD has yet been measured. The application of PECD in an analytical manner to such molecules remains challenging, and will require continued advancements in approaches for modeling PECD, in particular in terms of treating multiplet structures, which might exhibit different PECD. 

Our study provides a valuable addition to the presently scant literature documenting PECD in organometallic systems. We note that the class of molecules X-HFC$_{3}$, where X represents any trivalent lanthanide ion, provides ample opportunity for additional experiment. A useful subsequent study could focus on a nearly closed-shell system, such as the 4f$^{1}$ Ce-HFC$_{3}$, for which electron correlation effects are minimal and current models might be more applicable. We note that the MO with lowest IE in this system should be Ce 4f-based, in contrast to the present Eu molecule, reflecting the gradual stabilization of the metal 4f electrons across the lanthanide series and providing a further rationale for targeting the Ce system. Moreover, we point out that X-HFC$_{3}$ molecules also provide appealing opportunities to study PECD's ability to resolve fine-structure features in PES, for example spin-orbit split states. Such application will require advances in state-of-the-art methods for modeling photoionization of relativistic electrons.\cite{zapata_b-spline_2024} Additional experiment with this class of molecules is certainly warranted, and will serve to provide useful benchmarks against which theory can be tested.

\section*{\label{sec:level1}Supplementary Information}

\section*{\label{sec:level1}Author Contributions}

\textbf{Viktoria Brandt}: Formal analysis (equal); Investigation (equal); Validation (lead); Visualization (equal); Writing - original draft (equal); Writing - review \& editing (equal).

\textbf{Michele Pugini}: Investigation (equal); Writing - review \& editing (equal).

\textbf{Nikolas Kaltsoyannis}: Methodology (equal); Formal analysis (equal); Visualization (equal); Writing - review \& editing (equal).

\textbf{Gustavo Garcia}: Methodology (equal); Investigation (equal); Software (lead); Writing - review \& editing (equal).

\textbf{Ivan Powis}: Conceptualization (equal); Methodology (equal); Formal analysis (equal); Supervision (equal); Visualization (equal); Writing - review \& editing (lead).

\textbf{Laurent Nahon}: Conceptualization (equal); Methodology (equal); Investigation (equal); Supervision (equal); Writing - review \& editing (equal).

\textbf{Dominik Stemer}: Conceptualization (lead); Formal analysis (equal); Investigation (equal); Project administration (lead); Supervision (lead); Visualization (equal); Writing - original draft (lead); Writing - review \& editing (equal).

%\subsection{\label{sec:level2}Second-level heading: Formatting}

%\subsubsection{\label{sec:level3}Third-level heading: Citations and Footnotes}

\begin{acknowledgments}
The authors acknowledge Synchrotron SOLEIL for enabling the described experiments through user proposal 20221157. V.B. acknowledges support from the International Max Planck Research School for Elementary Processes in Physical Chemistry. N.K. thanks the University of Manchester for access to its Computational Shared Facility and associated support services. I.P. thanks the University of Nottingham for the provision of computational resources on the UoN HPC cluster. D.S. acknowledges funding from the European Research Council (Grant Agreement No. 883759). This work was supported by state funding from the ANR under the France 2030 program, with reference ANR-23-EXLU-0004, PEPR LUMA TORNADO. 

\end{acknowledgments}

\section*{Data Availability Statement}

\bibliography{PECD_Refs.bib}

\makeatletter\@input{xx2.tex}\makeatother

\end{document}

\title{Supporting Information for "Photoelectron Spectroscopy and Circular Dichroism of an Open-Shell Organometallic Camphor Complex"}

\date{}
\maketitle

\renewcommand{\thefigure}{S\arabic{figure}}
\renewcommand{\thetable}{S\arabic{table}}

\section*{Orbital and Ionization Energies}
\begin{table}[!ht]
    \caption{Calculated PBE0/cc-pVTZ HFC Conformer Energies}
    \centering
    \begin{tabular}{ccccc}
    \label{tab:confEn}
    \textbf{Structure}  & \textbf{E (a.u.)} & \textbf{$\Delta E$ (eV)} & \textbf{$\Delta E$ kj mol$^{-1}$} & \textbf{C$_3$ position}  \\ \hline
       Conf \# 9 & -1390.960382 & 0.000 & 0.0 & exo  \\
        Conf \# 1 & -1390.959604 & 0.021 & 2.0 & exo  \\ 
        Conf \# 5 & -1390.959142 & 0.034 & 3.3 & exo  \\ 
        Conf \# 3 & -1390.959054 & 0.036 & 3.5 & endo  \\ 
        Conf \# 6 & -1390.958328 & 0.056 & 5.4 & exo  \\ 
        Conf \# 7 & -1390.958173 & 0.060 & 5.8 & endo  \\ 
        Conf \# 4 & -1390.95611 & 0.116 & 11.2 & endo  \\ 
        Conf \# 8 & -1390.955986 & 0.120 & 11.5 & endo  \\ 
        Conf \# 10 & -1390.955654 & 0.129 & 12.4 & endo  \\ 
        Conf \# 2 & -1390.955195 & 0.141 & 13.6 & exo  \\  
        ~ & ~ & ~ & ~ & ~ \\
      	Enol \#1 & -1390.972407 & -0.327 & -31.6 & ~  \\ 
      Enol \#9 & -1390.971682 & -0.307 & -29.7 & ~  \\ \hline
    \end{tabular}
\end{table}

\begin{table}[!ht]
     \caption{OVGF/cc-pVTZ//PBE0/cc-pVTZ Calculated HFC Ionization Energies (eV)} 
    \centering
    \begin{tabular}{cccccc}
    \label{tab:confIE}
        \textbf{Structure} & \textbf{HOMO} & \textbf{HOMO-1} & \textbf{HOMO-2} & \textbf{HOMO-3} & \textbf{HOMO-4} \\ \hline
        Conf \#1 & 9.31 & 10.61 & 10.73 & 11.31 & 11.57 \\ 
        Conf \#2 & 9.32 & 10.58 & 10.69 & 11.28 & 11.58 \\ 
        Conf \#3 & 9.39 & 10.64 & 10.59 & 11.46 & 11.70 \\ 
        Conf \#4 & 9.30 & 10.71 & 10.64 & 11.38 & 11.79 \\ 
        Conf \#5 & 9.35 & 10.56 & 10.75 & 11.35 & 11.58 \\ 
        Conf \#6 & 9.34 & 10.58 & 10.75 & 11.33 & 11.57 \\ 
        Conf \#7 & 9.38 & 10.66 & 10.66 & 11.45 & 11.76 \\ 
        Conf \#8 & 9.33 & 10.71 & 10.57 & 11.42 & 11.62 \\ 
        Conf \#9 & 9.33 & 10.56 & 10.76 & 11.30 & 11.55 \\ 
        Conf \#10 & 9.31 & 10.72 & 10.55 & 11.40 & 11.70 \\ 
        ~ & ~ & ~ & ~ & ~ & ~ \\ 
        Enol \#1 & 8.88 & 9.69 & 10.78 & 11.40 & 11.74 \\ 
        Enol \#9 & 8.85 & 9.68 & 10.79 & 11.39 & 11.73 \\ \hline
    \end{tabular}
\end{table}

\begin{table}[!ht]
    \caption{ADC(3[4+])/cc-pVDZ//PBE0/cc-pVTZ Calculated HFC Ionization Energies (eV)}
    \centering
    \begin{tabular}{cccccc}
    \label{tab:IP-ADC}
        \textbf{Structure} & \textbf{HOMO} & \textbf{HOMO-1} & \textbf{HOMO-2} & \textbf{HOMO-3} & \textbf{HOMO-4} \\ \hline
        Conf \#1 & 9.23 & 10.34 & 10.96 & 11.41 & 11.73 \\ 
        Conf \#9 & 9.26 & 10.31 & 10.95 & 11.39 & 11.70 \\ 
        & & & & & \\
        Enol \#1 & 8.94 & 9.61 & 10.91 & 11.47 & 11.86 \\ 
        Enol \#9 & 8.88 & 9.59 & 10.90 & 11.44 & 11.83 \\ \hline
    \end{tabular}
    \label{tab:ADC-IP}
\end{table}
\newpage
\begin{figure}[!ht]
    \centering
    \includegraphics[width=0.65\linewidth]{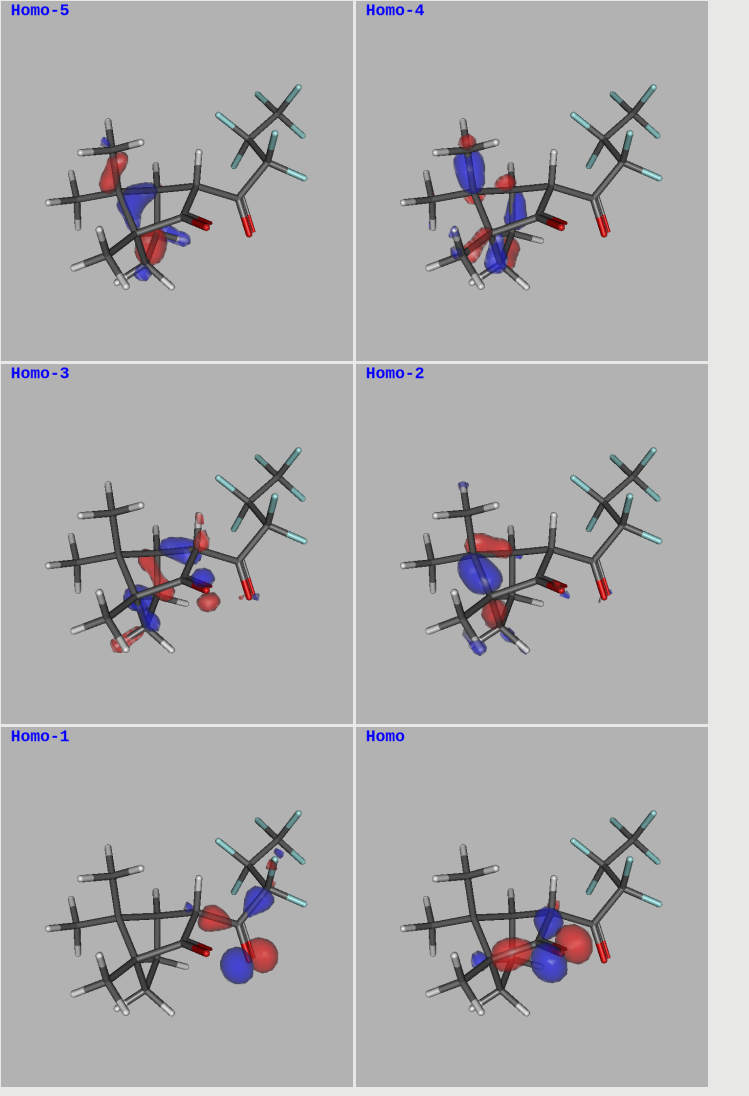}
    \caption{The outer six orbitals of HFC (Conformer \#9), calculated using the PBE0 hybrid functional with a cc-pVTZ basis set, corresponding to Table~\ref{tab:confIE}.}
    \label{fig:HFC6mos}
\end{figure}

\newpage

%IP-ADC(3;4+) / cc-pVDZ Calculation  [\textcolor{red}{Conformer 1 only}]\\
% %       PFBC\_IP-ADC3\_VDZ\_x10.out\\
% %       Basis:  Symm:  SubGroup:\\
% %       Geom Hash Code:  20FEA\\
%%        Host = hmcomp005.int.ada.nottingham.ac.uk\\
%        Q-Chem version 5.4.2\\

%\rowcolors{1}{white}{mygray}
%\begin{table}[h]
%\caption{Calculated ionization energies for the ten highest occupied molecular orbitals of HFC, in both the keto and enol form.} \label{tab:VIE_HFC-keto}
%\begin{center}
%\begin{tabular}{ccc}
%\#&IE$_{keto}$ (eV)&IE$_{enol}$ (eV)\\
%\hline
%1&9.288&\textcolor{red}{X}    \\
%2&10.458&\textcolor{red}{X}   \\
%3&10.961&\textcolor{red}{X}   \\
%4&11.399&\textcolor{red}{X}   \\
%5&11.707&\textcolor{red}{X}   \\
%6&12.041&\textcolor{red}{X}   \\
%7&12.372&\textcolor{red}{X}   \\
%8&12.485&\textcolor{red}{X}   \\
%9&12.787&\textcolor{red}{X}   \\
%10&13.023&\textcolor{red}{X}  \\
%\hline
%\end{tabular}
%\end{center}
%\end{table}

\noindent
%The first vertical ionization energy of Eu-HFC$_{3}$, calculated by the so-called $\Delta$-SCF method, is 7.695~eV. For HFC, in the keto (conformer \#9) and the enol (conformer \#1) form, the first ionization energies calculated using the same method are 8.949~eV and 8.894~eV, respectively.

\rowcolors{1}{white}{mygray}
\begin{table}[h]
\caption{Orbital energies for the neutral HFC and Eu-HFC$_{3}$ molecules.} 
\label{tab:Orbital_Energies}
\begin{center}
\begin{tabular}{cccc}
\#&HFC keto conformer \#9 (eV) & HFC enol conformer \#1 (eV) & Eu-HFC$_{3}$ (eV)  \\
\hline
1 & -7.251 & -7.254 &-6.689 \\
2 & -8.145 & -7.574 &-6.690 \\
3 & -8.902 & -8.818 &-6.707 \\
4 & -9.297 & -9.373 &-7.633 \\
5 & -9.616 & -9.694 &-7.633 \\
6 & -9.845 & -9.755 &-7.650 \\
7 & -10.053 & -10.040 &-8.500 \\
8 & -10.200 & -10.143 &-8.501 \\
9 & -10.583 & -10.211 &-8.546 \\
10 & -10.749 & -10.732 &-8.608 \\
11 & ~ & ~ &-8.608 \\
12 & ~ & ~ &-8.612 \\
\hline
\end{tabular}
\end{center}
\end{table}

\section*{Cartesian Coordinates (in \AA)}

\textbf{(1R,4R)-3-(Heptafluorobutyryl)-(+)-camphor, HFC}
\\
\\

\rowcolors{1}{white}{mygray}
\begin{longtable}{m{2cm}ccc}
\caption{Cartesian coordinates (in \AA) for optimized geometry of (1R,4R)-3-(heptafluorobutyryl)-(+)-camphor, HFC, in the keto form. These coordinates correspond to keto conformer \#9 in Table~\ref{tab:confEn}. The associated SCF energy is -1390.960382~H.} 

\label{tab:Ligand_Coordinates} \\

\hline
\endfirsthead

 \hline 
\endhead

 \hline
\endfoot

\endlastfoot

F & -2.366209 & -1.904265 & 0.379894 \\
F & -1.586242 & -1.181633 & -1.523515 \\
F & -1.626658 & 1.309841 & -0.734962 \\
F & -1.891310 & 0.760539 & 1.359450 \\
F & -4.360442 & -0.135735 & 0.878480 \\
F & -4.100446 & 0.066867 & -1.256747 \\
F & -4.074276 & 1.813364 & 0.010188 \\
O & 2.321444 & -2.222110 & -1.015874 \\
O & 0.048078 & -1.838604 & 1.289354 \\
C & 2.813704 & 1.140632 & -0.321288 \\
C & 3.344602 & -0.272551 & 0.039718 \\
C & 1.444424 & 0.930984 & 0.377862 \\
C & 0.924732 & -0.333602 & -0.350813 \\
C & 3.096742 & -0.335441 & 1.570817 \\
C & 1.853501 & 0.548152 & 1.799950 \\
C & 2.234172 & -1.136204 & -0.520844 \\
C & 3.623572 & 2.275742 & 0.289186 \\
C & 2.727329 & 1.388173 & -1.823756 \\
C & 4.721808 & -0.678877 & -0.407464 \\
C & -0.137351 & -1.115268 & 0.353462 \\
C & -1.591355 & -0.978665 & -0.189533 \\
C & -2.193201 & 0.417506 & 0.095481 \\
C & -3.729260 & 0.533238 & -0.072045 \\
H & 0.760963 & 1.776044 & 0.309619 \\
H & 0.565999 & -0.064056 & -1.344134 \\
H & 3.977164 & 0.045941 & 2.090366 \\
H & 2.943321 & -1.361869 & 1.905714 \\
H & 1.067898 & 0.024286 & 2.339312 \\
H & 2.092071 & 1.442038 & 2.375113 \\
H & 3.798635 & 2.160970 & 1.357083 \\
H & 4.597110 & 2.347373 & -0.199559 \\
H & 3.109719 & 3.227748 & 0.134755 \\
H & 3.726032 & 1.546995 & -2.234439 \\
H & 2.284711 & 0.563033 & -2.383065 \\
H & 2.142254 & 2.287159 & -2.031026 \\
H & 4.914985 & -1.718781 & -0.140103 \\
H & 4.828050 & -0.598750 & -1.490710 \\
H & 5.484624 & -0.056144 & 0.063737 \\

\hline
\end{longtable}

\rowcolors{1}{white}{mygray}
\begin{longtable}{m{2cm}ccc}
\caption{Cartesian coordinates (in \AA) for optimized geometry of (1R,4R)-3-(heptafluorobutyryl)-(+)-camphor, HFC, in the enol form. These coordinates correspond to enol conformer \#1 in Table~\ref{tab:confEn}. The associated SCF energy is -1390.972407~H.} \\

 \hline
\endfirsthead

 \hline 
\endhead

 \hline
\endfoot

\endlastfoot

C & -1.871088 & 0.854923 & -0.931431 \\
C & -3.090591 & 0.173425 & -0.356202 \\
C & -2.416029 & -1.072940 & 0.293056 \\
C & -1.385215 & -0.237302 & 1.100811 \\
C & -0.780322 & 0.601691 & 0.008925 \\
C & -2.291319 & 0.717897 & 1.900875 \\
C & -3.416156 & 1.041369 & 0.892354 \\
C & -3.353465 & -1.892593 & 1.165447 \\
C & -1.771692 & -2.004815 & -0.724452 \\
C & -4.225989 & -0.020122 & -1.322929 \\
C & 0.423665 & 1.150461 & -0.238390 \\
C & 1.641193 & 1.017332 & 0.647129 \\
C & 2.785973 & 0.131687 & 0.074295 \\
C & 2.368261 & -1.280985 & -0.398339 \\
F & 3.451867 & -1.987368 & -0.684672 \\
O & -1.795566 & 1.515414 & -1.961316 \\
O & 0.673830 & 1.874618 & -1.318480 \\
F & 1.300301 & 0.510939 & 1.846015 \\
F & 2.181947 & 2.232764 & 0.852715 \\
F & 3.707171 & -0.019009 & 1.036914 \\
F & 3.347482 & 0.753657 & -0.966642 \\
F & 1.614222 & -1.203158 & -1.489067 \\
F & 1.684359 & -1.917146 & 0.548815 \\
H & -0.686393 & -0.815715 & 1.697649 \\
H & -4.399404 & 0.760607 & 1.272706 \\
H & -3.466521 & 2.100310 & 0.635231 \\
H & -1.745553 & 1.603777 & 2.222609 \\
H & -2.680761 & 0.231778 & 2.795753 \\
H & -3.900330 & -1.299847 & 1.896766 \\
H & -4.086235 & -2.414729 & 0.546704 \\
H & -2.786347 & -2.651864 & 1.709337 \\
H & -2.535642 & -2.505382 & -1.322837 \\
H & -1.097970 & -1.487534 & -1.408644 \\
H & -1.192056 & -2.778171 & -0.215835 \\
H & -4.543187 & 0.938779 & -1.735987 \\
H & -3.929566 & -0.651868 & -2.161943 \\
H & -5.084109 & -0.479940 & -0.829327 \\
H & -0.181523 & 1.913276 & -1.827406 \\

\hline
\end{longtable}

\noindent
\textbf{Europium(III) tris[3-(heptafluoropropylhydroxymethylene)-(1R,4R)-camphorate], Eu-HFC$_{3}$}
\\
\\

\rowcolors{1}{white}{mygray}
\begin{longtable}{m{2cm}ccc}
\caption{Cartesian coordinates (in \AA) for optimized geometry of Europium(III) tris[3-(heptafluoropropylhydroxymethylene)-(1R,4R)-camphorate], Eu-HFC$_{3}$. The associated SCF energy is -4881.7388062~H.} \label{tab:Complex_Coordinates} \\

 \hline
\endfirsthead

%\multicolumn{3}{c}%
%{{\bfseries \tablename\ \thetable{} -- continued from previous page}} \\
% \multicolumn{1}{c|}{\textbf{First column}} & \multicolumn{1}{c|}{\textbf{Second column}} & \multicolumn{1}{c}{\textbf{Third column}} \\ 
 \hline 
\endhead

% \multicolumn{3}{|r|}{{Continued on next page}} \\ 
 \hline
\endfoot

\endlastfoot

Eu              & 0.000483  & -0.001882 & 1.776481  \\
O               & -1.543424 & -0.970947 & 0.400569  \\
O               & -2.030890 & 0.735517  & 2.617622  \\
O               & 0.376516  & -2.131875 & 2.612192  \\
O               & 1.658229  & 1.385257  & 2.615395  \\
O               & 1.609986  & -0.852553 & 0.397648  \\
O               & -0.067290 & 1.822046  & 0.403866  \\
C               & 2.284122  & 2.289376  & 2.014136  \\
C               & 0.825148  & 2.658764  & 0.073785  \\
C               & -2.714342 & -0.615851 & 0.071054  \\
C               & -3.126773 & 0.825344  & 2.016329  \\
C               & 0.847018  & -3.124644 & 2.009112  \\
C               & 1.887761  & -2.043355 & 0.065444  \\
C               & 1.559800  & -3.170993 & 0.780206  \\
C               & 1.967342  & 2.934072  & 0.787654  \\
C               & -3.525427 & 0.231674  & 0.787909  \\
C               & -3.269454 & -1.224080 & -1.226677 \\
C               & 2.692970  & -2.216276 & -1.232144 \\
C               & 0.571791  & 3.447590  & -1.220829 \\
C               & -2.246424 & -2.086420 & -2.006247 \\
C               & 2.928316  & -0.896511 & -2.007394 \\
C               & -0.688283 & 2.994174  & -1.998546 \\
C               & -2.748139 & -2.644788 & -3.365941 \\
C               & 3.663443  & -1.046861 & -3.367247 \\
C               & -0.926470 & 3.712585  & -3.354741 \\
F               & -3.696686 & -0.239410 & -2.046148 \\
F               & 2.055136  & -3.076824 & -2.054567 \\
F               & 1.636375  & 3.328777  & -2.043122 \\
F               & -4.335207 & -2.006891 & -0.939912 \\
F               & 3.903885  & -2.747586 & -0.945557 \\
F               & 0.426161  & 4.760948  & -0.929944 \\
F               & -1.154369 & -1.361607 & -2.272874 \\
F               & 1.754626  & -0.312199 & -2.272118 \\
F               & -0.605754 & 1.687115  & -2.270230 \\
F               & -1.915162 & -3.149579 & -1.255536 \\
F               & 3.682957  & -0.080494 & -1.253612 \\
F               & -1.772963 & 3.234652  & -1.244057 \\
F               & -2.995934 & -1.667221 & -4.224832 \\
F               & 2.941647  & -1.747132 & -4.229436 \\
F               & 0.044019  & 3.447227  & -4.216464 \\
F               & -3.846202 & -3.374008 & -3.213822 \\
F               & 4.844005  & -1.633482 & -3.216281 \\
F               & -1.015614 & 5.026946  & -3.196558 \\
F               & -1.795727 & -3.420753 & -3.874161 \\
F               & 3.859455  & 0.167826  & -3.870853 \\
F               & -2.073342 & 3.272231  & -3.862965 \\
C               & 3.106759  & 3.891059  & 0.525861  \\
H               & 2.889151  & 4.702024  & -0.165076 \\
C               & -4.925372 & 0.737132  & 0.528115  \\
H               & -5.517491 & 0.143803  & -0.164563 \\
C               & 1.823711  & -4.635587 & 0.518744  \\
H               & 2.633984  & -4.850145 & -0.173984 \\
C               & 3.600147  & 2.862611  & 2.481550  \\
C               & -4.283584 & 1.673465  & 2.486959  \\
C               & 0.691452  & -4.551257 & 2.477360  \\
C               & 0.461980  & -5.244433 & 0.134138  \\
C               & 4.310874  & 3.012130  & 0.137233  \\
C               & -4.770087 & 2.221071  & 0.144515  \\
C               & -0.333931 & -5.124461 & 1.450654  \\
C               & -4.267903 & 2.849029  & 1.461465  \\
C               & 4.605173  & 2.259874  & 1.451848  \\
C               & 3.465311  & 4.328153  & 1.969880  \\
C               & 2.028543  & -5.163169 & 1.962422  \\
C               & -5.482868 & 0.822182  & 1.972499  \\
C               & -4.234214 & 2.082921  & 3.933212  \\
H               & -4.252846 & 1.214770  & 4.594639  \\
H               & -3.316701 & 2.634574  & 4.144549  \\
H               & -5.081247 & 2.724537  & 4.182603  \\
C               & 3.934202  & 2.611458  & 3.926188  \\
H               & 3.195862  & 3.063297  & 4.591169  \\
H               & 3.949794  & 1.540581  & 4.135525  \\
H               & 4.915889  & 3.019972  & 4.172630  \\
C               & 0.311665  & -4.715782 & 3.923145  \\
H               & 1.072301  & -4.298422 & 4.585510  \\
H               & -0.625275 & -4.197967 & 4.134905  \\
H               & 0.180107  & -5.770655 & 4.170743  \\
C               & 2.089793  & -6.678955 & 2.068115  \\
H               & 2.188027  & -6.983074 & 3.112572  \\
H               & 1.215109  & -7.180560 & 1.658261  \\
H               & 2.967943  & -7.053495 & 1.536453  \\
C               & 3.262035  & -4.589162 & 2.646795  \\
H               & 3.292881  & -3.499631 & 2.616553  \\
H               & 3.301913  & -4.899854 & 3.693200  \\
H               & 4.167900  & -4.958374 & 2.160744  \\
C               & 2.355247  & 5.110927  & 2.658384  \\
H               & 1.394911  & 4.595291  & 2.629071  \\
H               & 2.607229  & 5.297691  & 3.704654  \\
H               & 2.223583  & 6.081495  & 2.174829  \\
C               & 4.750138  & 5.134968  & 2.075069  \\
H               & 4.967412  & 5.368700  & 3.119712  \\
H               & 5.619683  & 4.627198  & 1.661874  \\
H               & 4.636898  & 6.084530  & 1.546282  \\
C               & -5.600917 & -0.533744 & 2.655507  \\
H               & -4.672472 & -1.104614 & 2.622594  \\
H               & -5.887618 & -0.414063 & 3.702667  \\
H               & -6.374385 & -1.133476 & 2.170491  \\
C               & -6.826445 & 1.526404  & 2.080857  \\
H               & -7.137699 & 1.591332  & 3.125834  \\
H               & -6.825073 & 2.535511  & 1.672945  \\
H               & -7.590256 & 0.953715  & 1.549145  \\
H               & 0.002522  & -4.697278 & -0.685791 \\
H               & 0.566028  & -6.283584 & -0.180306 \\
H               & -0.690308 & -6.091947 & 1.808303  \\
H               & -1.205419 & -4.474395 & 1.360602  \\
H               & 4.474276  & 1.180668  & 1.360255  \\
H               & 5.622565  & 2.430874  & 1.807618  \\
H               & 4.062802  & 2.342673  & -0.683246 \\
H               & 5.160342  & 3.619184  & -0.178051 \\
H               & -4.927410 & 3.641094  & 1.820044  \\
H               & -3.269356 & 3.279072  & 1.371464  \\
H               & -4.066006 & 2.345206  & -0.675064 \\
H               & -5.721390 & 2.652078  & -0.169822 \\
\hline
\end{longtable}

\makeatletter\@input{xx.tex}\makeatother